# Proposed System for data hiding using Cryptography and Steganography

*Dipti Kapoor Sarmah[1], Neha Bajpai[2]

[1]*Department of Computer Engineering, Maharashtra Academy of Engineering,Pune, INDIA*
[2]*Department of Information Technology, Center of Development of advance computing,Noida, INDIA*
*Corresponding Author: e-mail: diptikap@rediffmail.com, Mob:9011073987*

**Abstract**

Steganography and Cryptography are two popular ways of sending vital information in a secret way. One hides the existence of the message and the other distorts the message itself. There are many cryptography techniques available; among them AES is one of the most powerful techniques. In Steganography we have various techniques in different domains like spatial domain, frequency domain etc. to hide the message. It is very difficult to detect hidden message in frequency domain and for this domain we use various transformations like DCT, FFT and Wavelets etc. In this project we are developing a system where we develop a new technique in which Cryptography and Steganography are used as integrated part along with newly developed enhanced security module. In Cryptography we are using AES algorithm to encrypt a message and a part of the message is hidden in DCT of an image; remaining part of the message is used to generate two secret keys which make this system highly secured.

***Keyword*:** Cryptography, Steganography, Stego- image, Threshold Value, DCT Coefficient

## 1. Introduction

Cryptography [1] and Steganography [1] are well known and widely used techniques that manipulate information (messages) in order to cipher or hide their existence respectively. Steganography is the art and science of communicating in a way which hides the existence of the communication. Cryptography scrambles a message so it cannot be understood; the Steganography hides the message so it cannot be seen. In this paper we will focus to develop one system, which uses both cryptography and Steganography for better confidentiality and security. Presently we have very secure methods for both cryptography and Steganography – AES algorithm is a very secure technique for cryptography and the Steganography methods, which use frequency domain, are highly secured. Even if we combine these techniques straight forwardly, there is a chance that the intruder may detect the original message. Therefore, our idea is to apply both of them together with more security levels and to get a very highly secured system for data hiding. This paper mainly focuses on to develop a new system with extra security features where a meaningful piece of text message can be hidden by combining security techniques like Cryptography and Steganography. As we know that-
• Hiding data is better than moving it shown and encrypted.
• To hide data in a popular object that will not attract any attention.
• In case the data is extracted, it will be encrypted.
But still there is a chance that the intruder can break the code. In our new system instead of applying existing techniques directly we will be using the following approach –
➢ Instead of hiding the complete encrypted text into an image, we will be hiding a part of the encrypted message.
➢ Unhidden part of the encrypted message will be converted into two secret keys.
➢ In this system to get the original message one should know, along with keys for Cryptography and Steganography, two extra keys and the reverse process of the key generation.

So our final goal of the project is to develop a new system which is highly secured and even if somebody retrieves the message from stego image [2] it becomes a meaningless for any existing cryptographic techniques.

## 2. Basic Concepts and Related work

There are many aspects to security and many applications. One essential aspect for secure communications is that of cryptography. But it is important to note that while cryptography is necessary for secure communications, it is not by itself sufficient. There are some specific security requirements [3] for cryptography, including Authentication, Privacy/confidentiality, and Integrity Non-repudiation. The three types of algorithms are described:
**(i) Secret Key Cryptography (SKC):** Uses a single key for both encryption and decryption
**(ii) Public Key Cryptography (PKC):** Uses one key for encryption and another for decryption
**(iii) Hash Functions:** Uses a mathematical transformation to irreversibly "encrypt" information.
Steganography is the other technique for secured communication. It encompasses methods of transmitting secret messages through innocuous cover carriers in such a manner that the very existence of the embedded messages is undetectable. Information can be hidden in images [5], audio, video, text, or some other digitally representative code. Steganography systems can be grouped by the type of covers [6] used (graphics, sound, text, executables) or by the techniques used to modify the covers
a) Substitution system [7].
b) Transform domain techniques [8]
c) Spread spectrum techniques [10]
d) Statistical method [11]
e) Distortion techniques [12]
f) Cover generation methods [12]

### 2.1 AES algorithm for Cryptography

This standard specifies the Rijndael algorithm [13], a symmetric block cipher that can process data blocks of 128 bits, using cipher keys with lengths of 128, 192, and 256 bits. The input, the output and the cipher key for Rijndael are each bit sequences containing 128, 192 or 256 bits with the constraint that the input and output sequences have the same length. In general the length of the input and output sequences can be any of the three allowed values but for the Advanced Encryption Standard (AES) the only length allowed is 128.

*2.1.1* **Advantages of using AES algorithm**

1. Very Secure.
2. Reasonable Cost.
3. Main Characteristics:
   I. Flexibility,   II. Simplicity

### 2.2 DCT [9]-frequency domain algorithm for Steganography

According to the method presented in this paper, the message is inserted into the DCT domain of the host image. The hidden message is a stream of "1" and "0" giving a total number of 56 bits. The transform is applied to the image as a multiple factor of 8x8 blocks. The next step of the technique after the DCT is to select the 56 larger positive coefficients, in the low-mid frequency range. The high frequency coefficients represent the image details and are vulnerable to most common image manipulation like filtering [14] compression [15] etc. Of course one might argue that this is the place where changes that come from watermarking [16] are more imperceptible, but this is true only if we're speaking of small sized blocks. Our scheme is applied to the whole image and since robustness is the main issue, the low and mid frequency coefficients are the most appropriate. The selected coefficients $c_i$ are ordered by magnitude and then modified by the corresponding bit in the message stream. If the $i$th message bit $s(i)$ to be embedded is "1", a quantity $D$ is added to the coefficient. This $D$ quantity represents the persistence factor. If the message bit is "0", the same quantity is subtracted from the coefficient. Thus the replaced DCT coefficients are
DCT (new) = DCT+1*D   for s(i)=1;
Else
DCT (new) = DCT-1*D   for s(i)=0.

DCT can separate the Image into High, Middle and Low Frequency components. To hide information we need to set a threshold value [17] for the DCT coefficients depending on the quality of the images.

### 2.2.1 Advantages of using frequency domain Steganography

- Very secure, hard to detect
- Flexible, different techniques for manipulation of DCT coefficients values

### 3. A Propose technique for combination

The design for the combining two different techniques is purely based on the idea – distort the message and hide the existence of the distorted message and for getting back the original message – retrieve the distorted message and regain the actual message by reversal of the distortion process.
Here we design the system with three modules-
- For Cryptography - Crypto Module
- For Steganography - Stego Module
- For extra security - Security Module

The extra security module that we are providing make this system highly secured. The process flow for the system is as follows-

### 3.1 Hiding the Text

- **Crypto Module** :

For Crypto Module the following steps are considered for encrypting the data (Refer **Figure1**):
➢ Insert text for encryption.
➢ Apply AES algorithm using 128 bit key (Key 1).
➢ Generate Cipher Text in hexadecimal form.

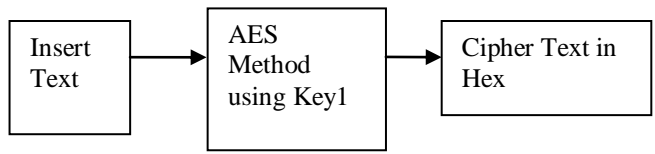

**Figure1:** Crypto Module

- **Security Module:**

This is an intermediate module which provides an extra security features to our newly developed system. This module is used to modify the cipher text and to generate two extra keys. In the reverse process it regenerates the original cipher text (Refer **Figure2**) .Before the hiding process this module works as follows:
➢ Separate the alphabets and digits from the cipher text.
➢ Keep track of the original position of the alphabet and the digits in the form of a secret key (Key 3).
➢ Separate first seven alphabets retrieved from first step and add the remaining alphabets at the end of the separated digits as in the first step. This generates the second key (Key 4).

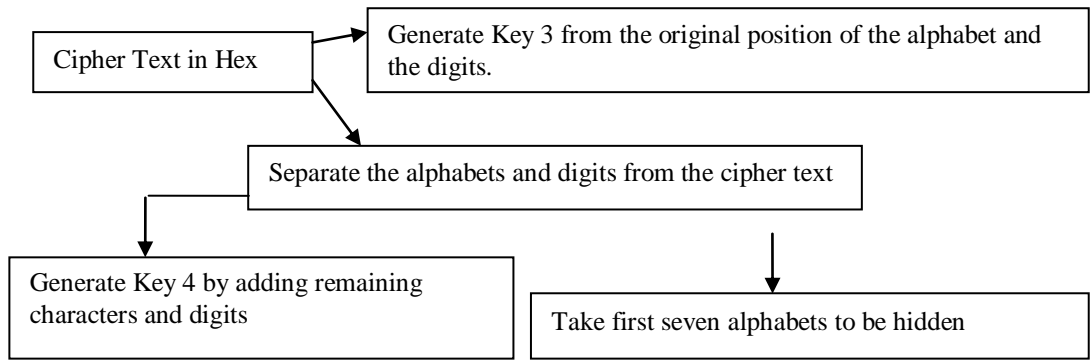

**Figure2:** Security Module

- **Stego Module:**

For Stego Module the following steps are considered for hiding the above generated Cipher text .For more details refer **Figure3**.
- ➢ Take seven alphabets from the above discussed Security Module.
- ➢ Scramble the alphabets using a 64 bit key (Key 2).
- ➢ Take a Gray Scale Image.
- ➢ Find the DCT of the Image.
- ➢ Hide the Cipher by altering DCTs.
- ➢ Apply Inverse DCT.
- ➢ Find the Stego Image.

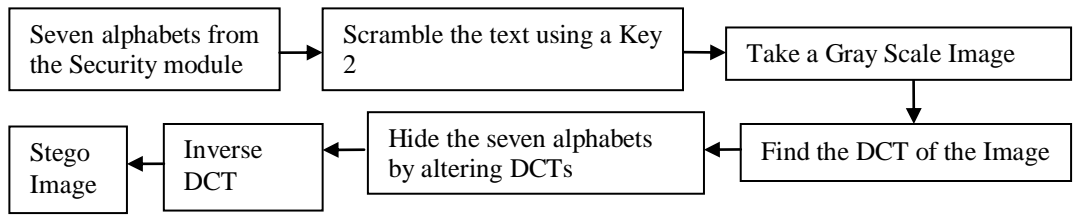

**Figure3:** Stego Module

### 3.2   Retrieving Text

- **Stego Module(Reverse Process) :**

For Stego Module the following steps are considered for retrieving the cipher text (Refer **Figure4**):
- ➢ Take DCT of the Original Image.
- ➢ Take DCT of the Stego Image.
- ➢ Take difference of DCT coefficients.
- ➢ Retrieve bits of the hidden seven alphabets from LSB of the DCT.
- ➢ Construct the distorted seven alphabets.
- ➢ Unscrambled the distorted seven alphabets using Key 2.
- ➢ Retrieve the original seven alphabets.

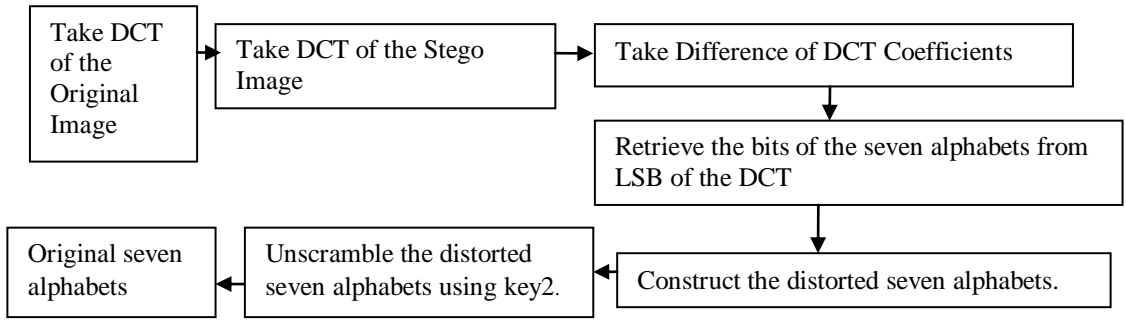

**Figure4:** Stego Module(Reverse Process)

- **Security Module(Reverse Process):**

For Security Module the following steps are considered for retrieving the cipher text (refer **Figure5**):
➢ Club the seven characters with the alphabets of Key 4.
➢ Using Key 3 and Key 4 reconstruct the cipher text from alphabets and digits.

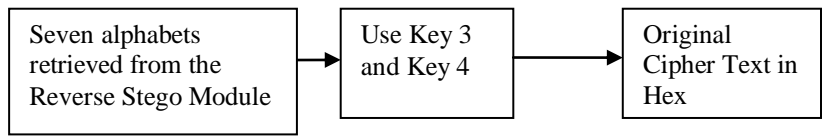

**Figure5:** Security Module (Reverse Process)

- **Crypto Module(Reverse Process):**

For Crypto Module the following steps are considered for retrieving the original text. For more details refer **Figure6:**
➢ Get the above retrieved cipher text.
➢ Reverse AES algorithm by using Key 1.
➢ Get the original message.

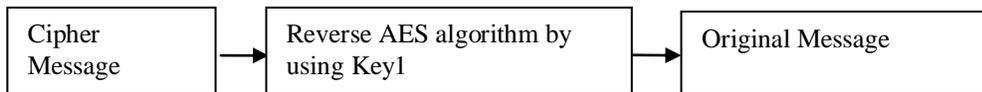

**Figure6:** Crypto Module (Reverse Process)

**4. Implementation Details**

This Project is mainly developed in VC6.0 plate-form using VC++. Here mainly three modules involved –
a)  Crypto Module - AES Implementation Module
b) Security Module – Newly developed technique
c)  Stego Module - DCT Techniques Implementation Module
These modules are design as reusable components and can work independently.

**4.1 Tools and Libraries used**

i.  Arisimage  Routines [18]
ii. Cximage599c [19]

These libraries are available with free license.

### *4.2* Algorithm for the proposed system

In AES Implementation, we are getting the cipher text in the hexadecimal form and the length of the cipher text is very large. In the newly proposed technique we partially hide the encrypted information in the image and with the help of the remaining part of the encrypted message we generate two keys. These two keys are secret keys and the receiver needs to know these two keys to retrieve the original encrypted message.
The steps for the algorithm are discussed below (Refer **Figure7 & Figure8**):

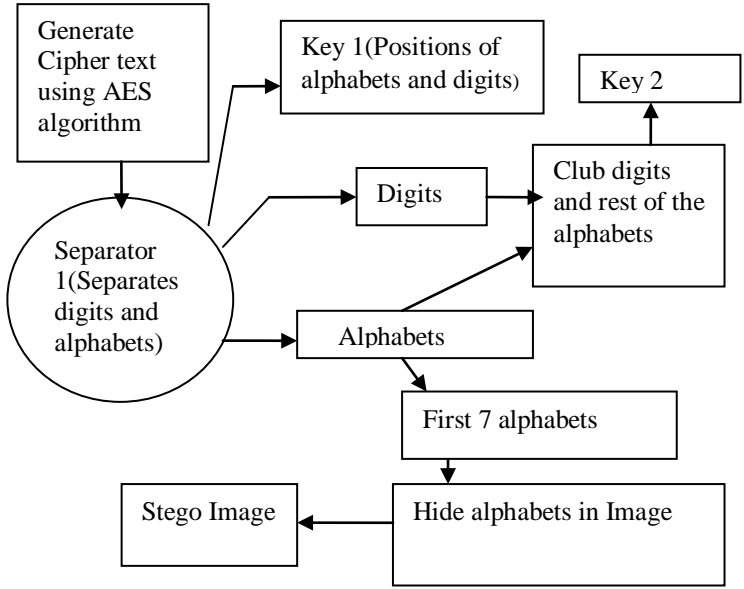

**Figure7:** Proposed System for hiding text

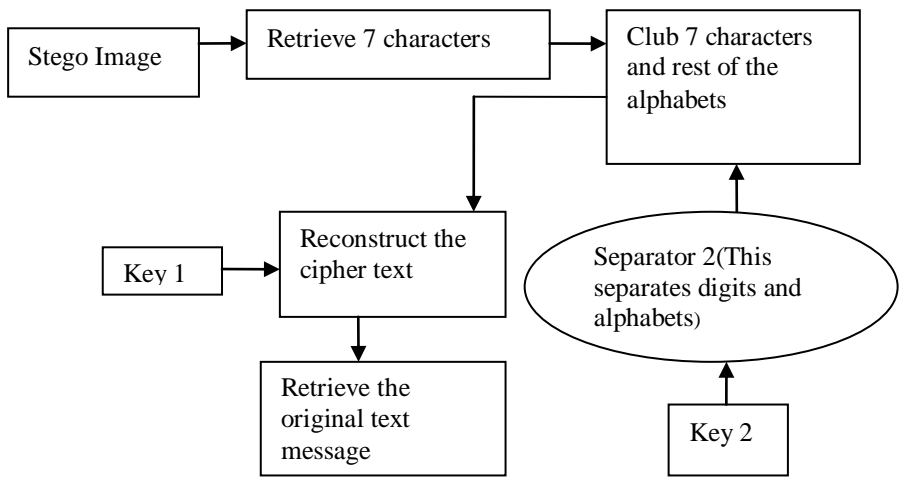

**Figure8:** Proposed System for retrieving text

### 4.2.1 Hiding Text

- Generate the cipher text in hexadecimal form by AES algorithm [13] in the form of alphabets (A, B, C, D, E, F) and digits (0, 1, 2, 3, 4, 5, 6, 7, 8, 9).
- Separate the alphabets and digits with the help of Separator 1 and keep track of the original position of the alphabets and digits in the form of the first key (**Key 1**).
- Take the first 7 characters of the alphabets; this part will be hidden in the image.
- Take the rest of the alphabets and combine with the digits; this will form the second key (**Key 2**).
- Hide the 7 characters in the Image as mentioned in 2.2.

### 4.2.2 Retrieving Text

- Retrieve the 7 characters from the image.
- Separate alphabets and digits from **Key 2** with the help of Separator 2.
- Add back the rest of the alphabets from **Key 2** to 7 characters retrieved from the image.
- Reorganize the alphabets and digits with the help of the **Key 1** to get back the original cipher text in hexadecimal form.
- Regenerate the original text message from the cipher text with the help of AES algorithm.

### 5. How secure the proposed system is?

The proposed solution is highly secure since-

❖ **It's a combination of two highly secured techniques -**
a) AES for cryptography
b) DCT manipulation for Steganography.

❖ **Number of Keys:** This system contains total 4 keys.
a) One 128 bits private key for AES algorithm
b) One 56 bits private key for scrambling the cipher text.
c) Two extra private generated keys for retrieving the original message.

These two extra generated keys make the system highly secured. If intruder detect the partial part of the hidden message from the stego image it will be totally meaningless for him and moreover until the complete set of keys are available getting the original message is impossible.

### 6. Conclusion

The work accomplished during this project can be summarized with the following points:
- In this project we have presented a new system for the combination of cryptography and Steganography using four keys which could be proven a highly secured method for data communication in near future.
- Steganography, especially combined with cryptography, is a powerful tool which enables people to communicate without possible eavesdroppers even knowing there is a form of communication in the first place. The proposed method provides acceptable image quality with very little distortion in the image.
- The main advantage of this Crypto/Stegno System is that the method used for encryption, AES, is very secure and the DCT transformation Steganography techniques are very hard to detect.

### 7. References

[1] Domenico Daniele Bloisi , Luca Iocchi: Image based Steganography and cryptography, Computer Vision theory and applications volume 1 , pp. 127-134 .
[2] Kharrazi, M., Sencar, H. T., and Memon, N. (2004). Image Steganography: Concepts and practice. In WSPC Lecture Notes Series
[3] D.R. Stinson, Cryptography: Theory and Practice, Boca Raton, CRC Press, 1995. ISBN: 0849385210
[4] Provos, N. and Honeyman, P. (2003). Hide and seek: An introduction to steganography. IEEE SECURITY & PRIVACY

[5] Chandramouli, R., Kharrazi, M. & Memon, N., "Image Steganography and steganalysis: Concepts and Practice", Proceedings of the 2nd International Workshop on Digital Watermarking, October 2003
[6] Owens, M., "A discussion of covert channels and steganography", SANS Institute, 2002
[7] Jamil, T., "Steganography: The art of hiding information is plain sight", IEEE Potentials, 18:01, 1999
[8] Stefan Katznbeisser, Fabien.A., P.Petitcolas editors, Information Hiding Techniques for Steganography and Digital Watermarking, Artech House, Boston. London, 2000.
[9] Wang, H & Wang, S, "Cyber warfare: Steganography vs. Steganalysis", Communications of the ACM, 47:10, October 2004
[10] Marvel, L.M., Boncelet Jr., C.G. & Retter, C., "Spread Spectrum Steganography", IEEE Transactions on image processing, 8:08, 1999
[11] Dunbar, B., "Steganography techniques and their use in an Open-Systems environment", SANS Institute, January 2002
[12] Bender, W., Gruhl, D., Morimoto, N. & Lu, A., "Techniques for data hiding", IBM Systems Journal, Vol 35, 1996
[13] C.E., Shannon, (1949), Communication theory of secrecy systems, Bell System Technical Journal, 28, 656-715.
[14] N. F. Johnson and S. Katzenbeisser, .A survey of steganographic techniques., in S. Katzenbeisser and F. Peticolas (Eds.): Information Hiding, pp.43-78. Artech House, Norwood, MA, 2000.
[15] Currie, D.L. & Irvine, C.E., "Surmounting the effects of lossy compression on Steganography", 19$^{th}$ National Information Systems Security Conference, 1996
[16] G., Derrick, (2001), Data watermarking Steganography and watermarking of digital data, Computer Law & Security Report, 17 (2), 101-104.
[17] Ross J. Anderson, Fabien A.P. Petitcolas, "On The Limits of Steganography", IEEE Journal of Selected Areas in Communications, 16(4): 474-481, May 1998. Special Issue on Copyright & Privacy Protection. ISSN 0733-8716.
[18] http://www.codeproject.com/KB/library/ArisFFTDFTLibrary.aspx
[19] http://www.xdp.it

**Biographical notes**

- **DIPTI KAPOOR SARMAH**, Assistant Professor (Computer Engineering Department) at Maharashtra Academy of Engineering received M.Tech in the year 2007 from Indraprastha University, Delhi. She has more than 7 years of teaching experience to graduate and postgraduate students in various engineering colleges. Proficient in the subjects related to Cryptography, DAA, C, C++, Data Structures, and Computer Graphics. She has completed various administrative responsibility such as aptitude test for students, Tutor In charge, Practical/Oral Examination In charge, NBA (National Board Accreditation) In charge, ISO In charge, Arrangement of Seminars of T.E. students and Class Teacher coordinator. She has published various research papers in various national and International conferences as well as attended various workshops and seminar. She is having Life Time membership of ISTE, IAENG(International association of engineers),IACSIT(International Association of Computer Science and Informationand ISSTE(International Society of Scientists, Technocrats and Engineers).

- **NEHA BAJPAI (Lecturer, School of IT)** recieved M.tech (Information Technology) from the Vinayaka Mission University of Tamilnadu in the year 2005. She has eight years of teaching and one year of IT implementation experience. Proficient in the subjects related to Object Oriented Analysis & Design, UML, Software Testing, Object Oriented Technologies (Java Programming) and Object Oriented Database Management System. She has worked on various projects in her Nine years of teaching and software implementation time span.
She also served, coordinated and taught various International Training Programs under Indo-Vietnam bi-lateral Cooperation and ITEC/SCAAP Scheme of MEA. She was an active participant and an organizing committee member for ASCNT 2009 held in CDAC, Noida.